\documentclass[
  aps,
  prmaterials,
  reprint,
  superscriptaddress,
  amsmath,
  amssymb,
  floatfix
]{revtex4-2}

\usepackage{amsmath,amssymb}
\usepackage{booktabs}
\usepackage{graphicx}
\usepackage{siunitx}
\usepackage{lmodern}
\usepackage{array}[=2016-10-06]

\newcommand{\sluschi}{\textsc{SLUSCHI}}
\newcommand{\sluschiup}{\textsc{SLUSCHI-UP}}
\newcommand{\umlp}{uMLIP}
\newcommand{\umlps}{uMLIPs}
\newcommand{\Tm}{T_{\mathrm{m}}}
\newcommand{\Pliq}{P_{\mathrm{liq}}}

\begin{document}

\title{\sluschiup: A Web Infrastructure for \sluschi{} Melting-Temperature Calculations Using Universal Machine-Learning Interatomic Potentials}

\author{Qi-Jun Hong}
\email{qhong@alumni.caltech.edu}
\affiliation{Department of Materials Science and Engineering, Arizona State University, Tempe, Arizona 85287, USA}

\date{\today}


\begin{abstract}
Melting temperature is a critical property for high-temperature materials design, but first-principles melting calculations based on finite-temperature molecular dynamics can require substantial computational resources. The \sluschi{} method reduces this cost by using small-cell solid--liquid coexistence simulations and statistical analysis of many short molecular-dynamics trajectories. Here I present \sluschiup{}, a deployed web service for atomistic melting-temperature estimation that couples the \sluschi{} workflow to selectable pretrained universal machine-learning interatomic potentials (uMLIPs) and asynchronous GPU execution. Users submit a crystal structure through a Materials Project identifier or POSCAR input, select a uMLIP backend, and launch a queued melting calculation without local installation of simulation software.
The current production interface supports \texttt{mace-mpa-0-medium}, \texttt{Allegro-OAM-L}, and \texttt{DPA-3.2-5M-OMat24}, while beta deployments expose additional models. On the compact MeltBench-10 validation set, the three production backends produce raw coexistence mean absolute errors in the range of approximately 178--327~K. Across the broader set of materials tested so far in MeltBench, the current deployed-job snapshot contains 119 raw uMLIP entries, and PBE-corrected Allegro-OAM-L predictions reach a mean absolute error of approximately 166~K. These values should be interpreted as screening-level infrastructure validation rather than a definitive ranking of uMLIPs. The results demonstrate that \sluschiup{} provides a practical, provenance-aware deployment layer between fast scalar melting-temperature predictors and much more expensive first-principles coexistence calculations, while retaining the usual limitations of uMLIP transferability, finite-size sampling, and high-temperature trajectory stability.
\end{abstract}


\maketitle

\section{Introduction}

Melting temperature, $\Tm$, is one of the most important thermodynamic properties for materials selection, phase-diagram construction, and the design of high-temperature materials. It constrains the use of refractory ceramics, turbine and hypersonic materials, metallurgical systems, additive-manufacturing feedstocks, molten salts, geophysical minerals, and multicomponent alloys. Despite its importance, $\Tm$ remains difficult to determine in high-throughput settings. Experiments can be slow, hazardous, or unavailable for unstable and multicomponent compounds, while first-principles calculations are computationally demanding because they require extensive finite-temperature sampling. Conventional approaches such as thermodynamic integration and large-cell solid--liquid coexistence simulations can consume on the order of $10^{4}$--$10^{5}$ CPU-hours per material, making routine screening impractical.

Several computational strategies have been developed to estimate melting temperatures. Free-energy methods compare the relative thermodynamic stability of solid and liquid phases, with representative approaches including particle-insertion chemical potentials, pair-distribution-function entropy, two-phase thermodynamics, thermodynamic integration, and generalized entropy calculations \cite{widom1963fluids,hong2012chemical,lin2003twophase,hong2025generalizedentropy}. Large-cell coexistence simulations directly evolve a solid--liquid interface in molecular dynamics and determine the melting temperature from interface motion or phase stability \cite{frenkel2001understanding,morris1994aluminum,alfe2001complementary,cazorla2007molybdenum}. The small-size coexistence approach provides a statistical alternative: instead of stabilizing a large interface, it launches many small coexistence simulations and infers the melting point from the probability that an initially mixed solid--liquid configuration evolves into a solid-like or liquid-like state \cite{hong2013small}. The corresponding \sluschi{} code automated this workflow for first-principles molecular dynamics, originally through VASP \cite{hong2016sluschi,kresse1996vasp}. Although \sluschi{} substantially reduces the cost of melting calculations relative to conventional large-interface approaches, its original density-functional-theory force evaluations can still require days to weeks of computation for a single material.

Recent advances in universal machine-learning potentials, here abbreviated as \umlps{}, make it possible to accelerate this workflow without constructing a system-specific empirical potential for each material. Models such as M3GNet \cite{chen2022m3gnet}, CHGNet \cite{deng2023chgnet}, MatterSim \cite{yang2024mattersim}, and MACE-MP \cite{batatia2024macefoundation} have demonstrated broad chemical coverage for atomistic energy and force prediction. In parallel, our recent work showed that coupling \sluschi{} to LAMMPS and pre-trained machine-learning potentials can reduce computational cost by at least an order of magnitude while retaining useful screening accuracy \cite{campbell2026sluschi,thompson2022lammps}. These developments create an opportunity to move melting-temperature calculations from expert-driven command-line workflows toward broadly accessible, atomistic, and reproducible web infrastructure.

Here I introduce \sluschiup{}, a deployed web service that couples the \sluschi{} melting workflow with selectable \umlps{}. Users provide a crystal structure through a Materials Project identifier or pasted POSCAR content, select a supported potential backend such as \texttt{mace-mpa-0-medium}, \texttt{Allegro-OAM-L} \cite{musaelian2023allegro,allegrooaml}, or \texttt{DPA-3.2-5M-OMat24} \cite{zhang2025dpa3,dpa325m}, and submit the job through an email-verified web interface. The service executes the required molecular dynamics on GPUs in the background, tracks metadata such as structures, model choices, checkpoints, and job identifiers, and returns a melting-temperature estimate with associated diagnostics. \sluschiup{} therefore fills the gap between fast formula-based predictors, which can return scalar estimates but lack atomistic coexistence information \cite{hong2022gnn}, and expensive first-principles coexistence calculations \cite{hong2013small,hong2016sluschi}. In this paper, I describe the deployed infrastructure, supported \umlp{} backends, melting workflow, and compact MeltBench-10 validation results \cite{meltbenchweb}, while emphasizing that predictions remain screening estimates whose accuracy depends on model transferability, finite-size sampling, and the stability of high-temperature solid--liquid trajectories.

The contribution of \sluschiup{} is not a new melting-temperature algorithm, but a deployed infrastructure layer that makes atomistic melting calculations broadly accessible. The platform integrates the established \sluschi{} small-size solid--liquid coexistence methodology with modern universal machine-learning interatomic potentials through a browser-based interface, eliminating the need for users to install \sluschi{}, LAMMPS, or model-specific software. A user submits a crystal structure through a Materials Project identifier or POSCAR file, selects a supported uMLIP backend, and the remaining workflow (including structure preparation, coexistence-cell construction, molecular-dynamics execution, trajectory analysis, and melting-temperature estimation) is handled automatically through an asynchronous GPU-backed service. To encourage community use and independent validation, the public deployment currently allows one free calculation per verified user every 24 hours. By combining automation, provenance tracking, reproducibility, and public accessibility, \sluschiup{} lowers the barrier to atomistic melting-temperature prediction and provides a common platform for benchmarking and validating universal machine-learning interatomic potentials. More broadly, the service has the potential to establish a new standard for melting-temperature calculations by making advanced coexistence simulations accessible, repeatable, and transparent to a much wider materials-science community.

\begin{figure}
\includegraphics[width=0.29\textwidth]{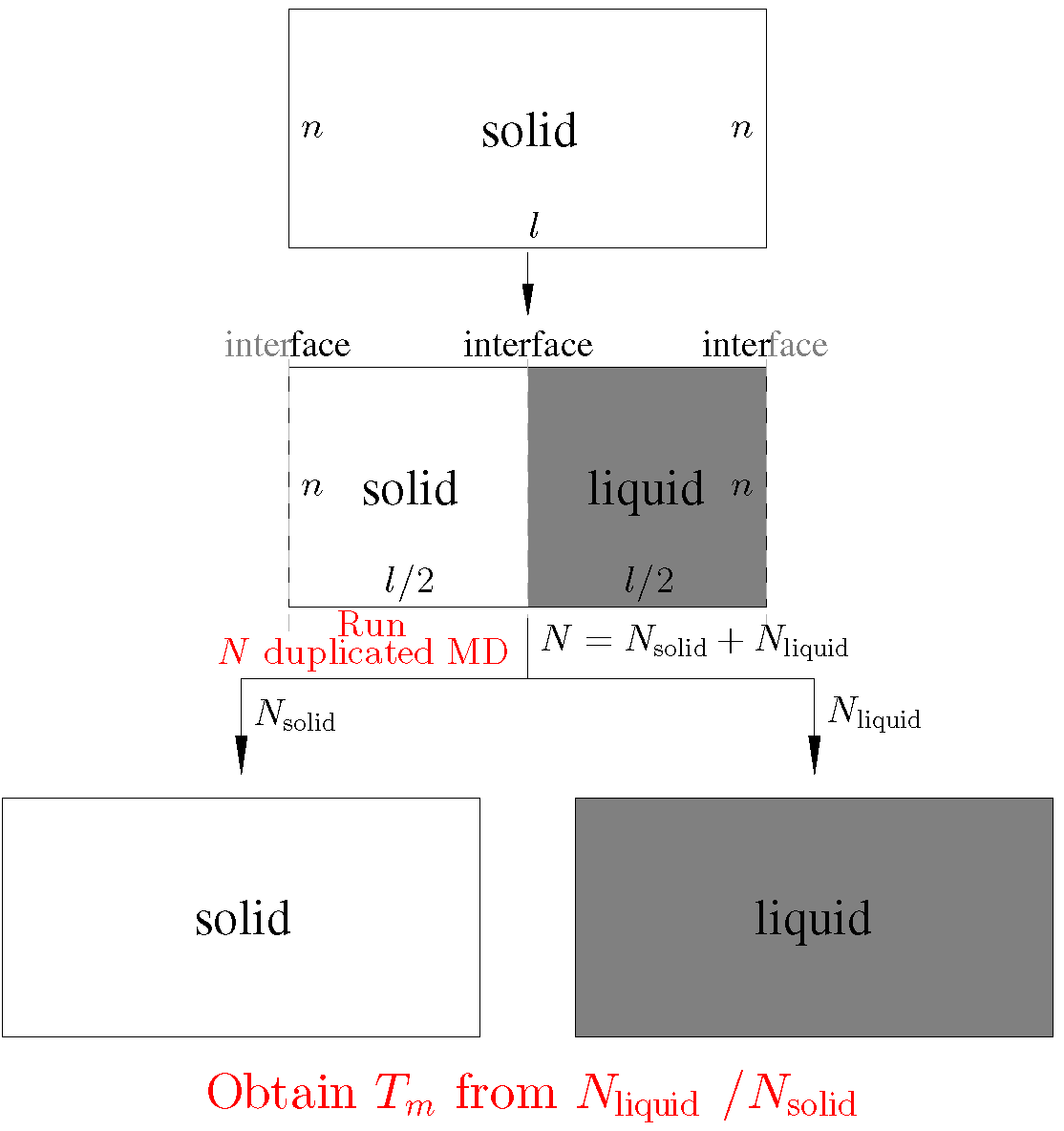}
\caption{Schematic of the SLUSCHI small-size coexistence method. An initially prepared solid–liquid coexistence cell is evolved through multiple independent molecular-dynamics trajectories at trial temperatures. Each trajectory is classified according to whether it evolves toward a solid-like or liquid-like state, and the melting temperature is inferred statistically from the balance between the two outcomes. Detailed descriptions of the methodology and implementation are available in Refs.~\cite{hong2013small,hong2016sluschi}.}
\label{fig:sluschi}
\end{figure}

\begin{figure}
\includegraphics[width=0.49\textwidth]{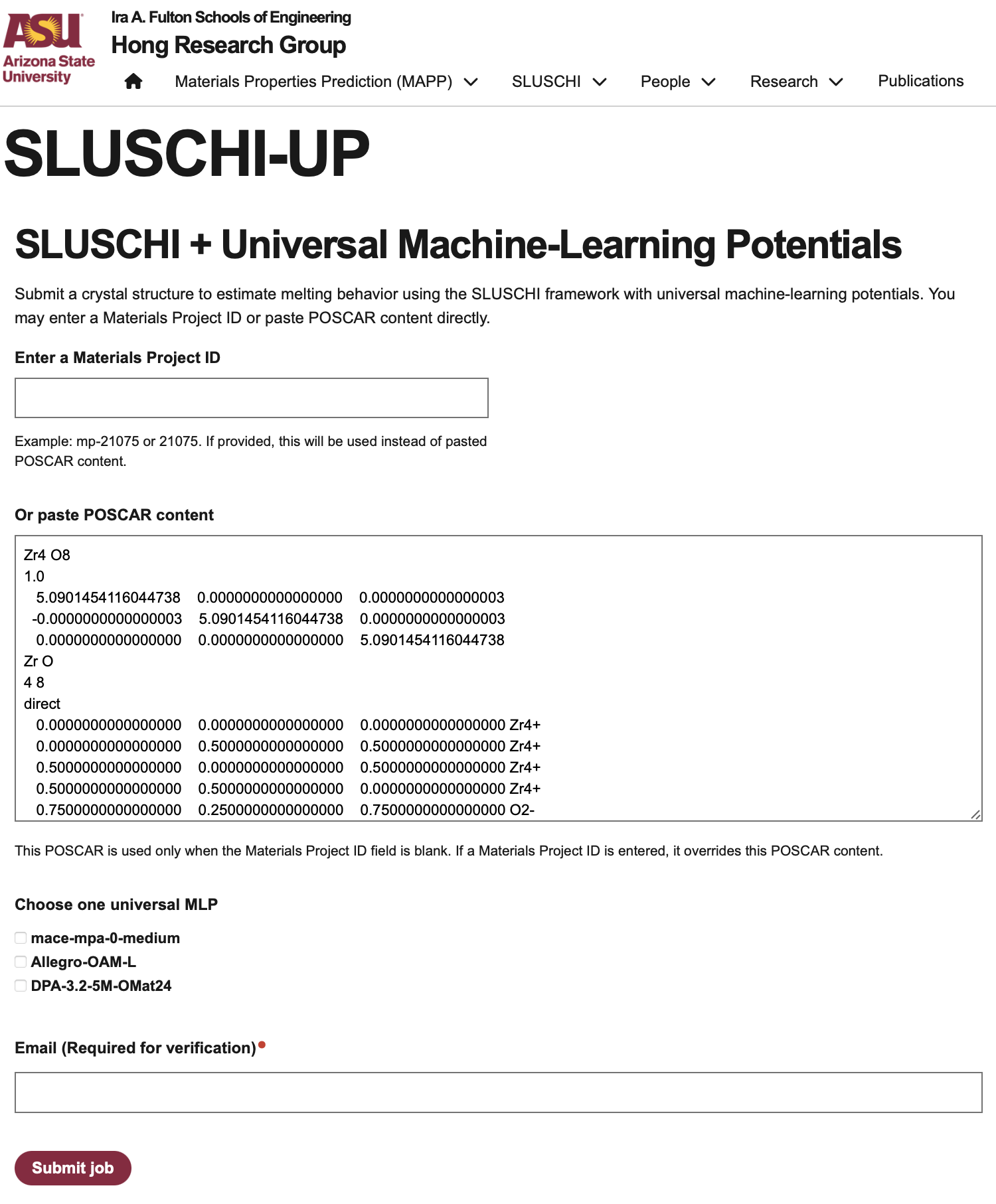}
\caption{Screenshot of the deployed \sluschiup{} web interface. The public service accepts a Materials Project identifier or POSCAR input, allows the user to select one supported universal machine-learning potential, and submits the calculation to an email-verified shared GPU queue for \sluschi{}-style melting-temperature estimation.}
\label{fig:sluschiup_interface}\label{fig:sluschiup_workflow}
\end{figure}

\newcommand{\tcell}[2]{\parbox[t]{#1}{\raggedright\strut #2\strut}}
\begin{table*}[t]
\centering
\small
\setlength{\tabcolsep}{4pt}
\caption{Production uMLIP backends exposed in the deployed \sluschiup{} interface and used in the MeltBench-10 validation. Version labels are the deployed checkpoint labels in the service.}
\label{tab:supported_umlips}
\begin{tabular}{@{}lllll@{}}
\toprule
Deployed label & Model family & Training/source data & Deployment path & Citation \\
\midrule
\texttt{mace-mpa-0-medium} &
MACE-MPA-0 &
MPTrj + sAlex / MACE foundation model &
ASE/MACE calculator &
\cite{batatia2024macefoundation,macefoundations} \\
\texttt{Allegro-OAM-L} &
Allegro / NequIP &
OMat24 pretraining + sAlex/MPTrj fine-tuning &
NequIP/Allegro, ASE/LAMMPS-compatible &
\cite{batzner2022nequip,musaelian2023allegro,allegrooaml} \\
\texttt{DPA-3.2-5M-OMat24} &
DPA-3 / DeePMD &
OpenLAM / OMat24 branch &
DeePMD-compatible backend &
\cite{zhang2025dpa3,dpa325m} \\
\bottomrule
\end{tabular}
\end{table*}

\section{Current Public Deployment}

\sluschiup{} builds on a broader history of materials cyberinfrastructure. Platforms such as the Materials Project and MPContribs provide web and API access to computed and contributed materials data \cite{jain2013materialsproject,huck2015mpcontribs}. OpenKIM provides standardized infrastructure for interatomic models, model metadata, testing, and reproducibility \cite{tadmor2011openkim,elliott2011kimapi}. AiiDAlab demonstrates how browser-based workflow interfaces can broaden access to automated atomistic calculations while preserving provenance through AiiDA \cite{yakutovich2020aiidalab}. \sluschiup{} is not intended to replace these general platforms; rather, it contributes a melting-specific deployment layer that combines \sluschi{} small-size coexistence, selectable pretrained uMLIP backends, asynchronous GPU execution, and job-level provenance for atomistic melting-temperature estimation.

\sluschiup{} is designed to make \sluschi{}-style melting-temperature calculations accessible through a web interface, without requiring users to install \sluschi{}, LAMMPS, a \umlp{} package, or a queue-system client locally. The deployed service accepts crystal-structure input through either a Materials Project \cite{jain2013materialsproject} identifier or pasted VASP POSCAR content, stores job metadata for provenance, and runs the requested melting calculation through an asynchronous GPU-backed workflow. In contrast to direct composition-to-\(\Tm\) predictors, \sluschiup{} preserves the atomistic coexistence workflow: the selected \umlp{} supplies energies and forces for molecular dynamics, while the melting temperature is inferred from the statistical outcome of solid--liquid coexistence trajectories.

The platform builds on the original \sluschi{} workflow, in which molecular dynamics trajectories were driven by first-principles calculations through VASP and the code automatically constructed a suitable supercell, prepared solid--liquid coexistence, and estimated the melting temperature from a user-provided crystal structure \cite{sluschigithub}. Figure~\ref{fig:sluschi} summarizes the underlying small-size coexistence methodology. Detailed descriptions of the coexistence-cell construction procedure, trajectory duplication strategy, statistical framework, convergence behavior, and first-principles implementation are available in the original SLUSCHI publications \cite{hong2013small,hong2016sluschi}. Although this approach is broadly transferable, it can require days to weeks and thousands to hundreds of thousands of CPU hours \cite{sluschigithub}. \sluschiup{} preserves the same statistical coexistence logic but replaces the electronic-structure force evaluation step with a selected \umlp{} backend. This design is motivated by our recent work showing that coupling \sluschi{} to LAMMPS and pre-trained machine-learning potentials can reduce computational cost by more than one order of magnitude relative to conventional DFT-based simulations on a benchmark set of 30 material systems \cite{campbell2026sluschi}.

The public deployment is hosted by the Hong Research Group at Arizona State University and is described on the website as ``\sluschi{} + Universal Machine-Learning Potentials.'' A user submits a structure, selects one supported \umlp{} backend, provides an email address for verification, and receives a job identifier that can be queried through the status interface. Because the calculations require multiple molecular dynamics trajectories and are GPU-intensive, jobs are processed through a shared queue; a typical turnaround time is approximately 12--24 hours \cite{sluschiupweb}. The production workflow separates the web request from the simulation tasks, allowing the interface to remain responsive while molecular dynamics and analysis jobs run asynchronously.

The current public interface supports two structure-submission modes. In the database-driven mode, the user enters a Materials Project identifier, such as \texttt{mp-21075} or \texttt{21075}. In the user-supplied mode, the user pastes POSCAR-format structural content directly into the web form. This design supports both standard database structures and structures generated from local calculations, collaborators, or external databases. When both a Materials Project identifier and POSCAR content are provided, the Materials Project identifier takes precedence. 

The current deployment exposes a single-choice model selector labeled ``Choose one universal Machine Learning Interactomic Potential.'' Each submitted job selects one backend from the currently supported public options: \texttt{mace-mpa-0-medium}, \texttt{Allegro-OAM-L}, and \texttt{DPA-3.2-5M-OMat24}. The one-model-per-job design simplifies queue accounting and user interpretation, while still allowing the same structure to be resubmitted with different \umlps{} for model-comparison studies. These three backends represent complementary families of modern universal machine-learning potentials: MACE uses higher-order equivariant message passing, Allegro uses a local equivariant architecture designed for scalable atomistic dynamics, and DPA-3 uses a line-graph-series message-passing framework designed for large atomistic models. In \sluschiup{}, the common requirement is not architectural uniformity, but the ability to provide stable energies and forces for finite-temperature molecular dynamics in chemically diverse materials. An expanded beta interface also exposes additional experimental backends, including \texttt{PET-MAD-S-v1.5}, \texttt{PET-OAM-XL-v1.0}, \texttt{SevenNet-Omni-i12}, and \texttt{CHGNet-0.3.0}; these beta models are not included in the MeltBench-10 statistics reported below. These additional beta backends broaden the architecture and training-data diversity of \sluschiup{}, including PET-MAD models trained for broad materials coverage, OMat24-derived models, SevenNet-Omni cross-domain transferable potentials, and CHGNet charge-informed graph neural-network potentials \cite{mazitov2025petmad,malosso2026mad15,barroso2024omat24,kim2025sevennetomni,deng2023chgnet}. 

\section{Melting-Temperature Workflow}

The \sluschiup{} workflow begins with a crystal structure supplied either as a Materials Project identifier or as pasted POSCAR content. The structure is parsed, checked for element validity, standardized, and converted into a form suitable for coexistence-cell construction. When a Materials Project identifier is provided, the corresponding database structure is used as the authoritative input. This structure-based workflow is important because melting behavior depends on the selected polymorph and initial crystal structure, not only on chemical composition.

\begin{figure}
\includegraphics[width=0.49\textwidth]{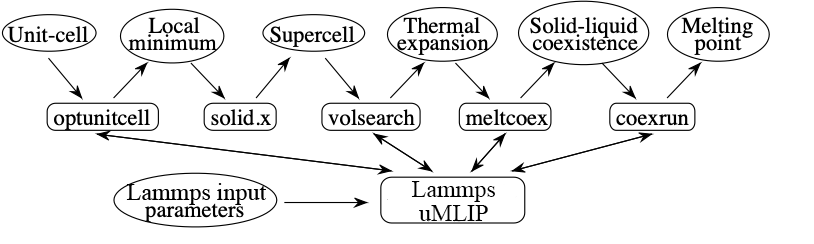}
\caption{Schematic of the SLUSCHI-UP melting temperature calculation workflow. \label{workflow}}
\end{figure}

After input normalization, the user-selected \umlp{} backend is checked for elemental coverage and availability. In the current deployment, each submitted job uses one backend selected from the public model choices. As shown in Fig. \ref{workflow}, the selected \umlp{} supplies the energies and forces required for molecular dynamics, replacing the electronic-structure force evaluations used in the original VASP-based \sluschi{} workflow. Model coverage does not guarantee accuracy for all high-temperature liquid or interfacial configurations, so the workflow treats the selected potential as part of the job provenance and not as an interchangeable black box.

Given the normalized solid structure, \sluschiup{} constructs a small solid--liquid coexistence cell following the logic of \sluschi{} \cite{hong2016sluschi}. One half portion of the cell is prepared as a liquid-like region, while the remaining half portion is retained as solid-like. Independent molecular dynamics trajectories are then launched at trial temperatures. Let $T_i$ be a trial temperature and let $r=1,\ldots,n_i$ index the independent replicas. Each trajectory is classified according to whether it evolves toward a liquid-like or solid-like state:
\begin{equation}
y_{ir} =
\begin{cases}
1, & \text{trajectory ends liquid-like},\\
0, & \text{trajectory ends solid-like}.
\end{cases}
\end{equation}
The empirical liquid probability at $T_i$ is then
\begin{equation}
\widehat{\Pliq}(T_i)=\frac{1}{n_i}\sum_{r=1}^{n_i} y_{ir}.
\end{equation}

The melting temperature is estimated from the transition region where the probability of a liquid-like outcome crosses one half. A simple parametric representation is
\begin{equation}
\Pliq(T)=\frac{1}{1+\exp[-(T-\Tm)/w]},
\end{equation}
where $\Tm$ is the melting temperature and $w$ is a finite-size transition-width parameter. If $k_i$ out of $n_i$ replicas end liquid-like at temperature $T_i$, the likelihood is
\begin{equation}
\mathcal{L}(\Tm,w)
=
\prod_i
\binom{n_i}{k_i}
\Pliq(T_i)^{k_i}
\left[1-\Pliq(T_i)\right]^{n_i-k_i}.
\end{equation}
The reported melting temperature is the fitted value at which $\Pliq(T)=0.5$, and the statistical uncertainty is obtained from the fitted transition region or from resampling of the replica outcomes.
Detailed descriptions of the coexistence-cell construction procedure, trajectory duplication strategy, statistical framework, convergence behavior, and first-principles implementation are available in the original SLUSCHI publications \cite{hong2013small,hong2016sluschi}.


Each completed job stores the input structure, selected \umlp{} backend, job identifier, trajectory-level outcomes, fitted melting-temperature estimate, reported uncertainty, and relevant warnings or failure information. Failed or inconclusive calculations are recorded explicitly because they provide useful information about workflow robustness, model transferability, and the limits of uMLIP-driven melting simulations.

\section{Results: MeltBench-10 Validation}

\subsection{Validation scope and metrics}

I evaluate \sluschiup{} using the compact MeltBench-10 set. This set is used here as an infrastructure-validation dataset rather than as a complete benchmark of universal machine-learning potentials. Its purpose is to demonstrate that the deployed service can accept structurally diverse materials, run \sluschi{}-style coexistence calculations with selectable \umlps{}, and return quantitative melting-temperature estimates with traceable job metadata. A systematic comparison of individual \umlp{} backends, chemistry-class dependence, correction schemes, finite-size effects, and uncertainty calibration is reserved for the companion MeltBench study.

For each material, I compare the \sluschiup{} melting-temperature estimate, $T_m^{\rm SU}$, against the reconciled reference melting temperature, $T_m^{\rm ref}$. The signed error is defined as
\begin{equation}
\Delta T_m = T_m^{\rm SU} - T_m^{\rm ref}.
\end{equation}
I report the signed error, absolute error, mean absolute error, root-mean-square error, median absolute error, and the fraction of predictions within 200~K of the reference value.

For entries where it is available, MeltBench also reports a first-order PBE enthalpy correction following the correction strategy used in prior \sluschi{}+machine-learning-potential work \cite{campbell2026sluschi}. The correction is based on the thermodynamic relation
\begin{equation}
\Tm = \frac{\Delta H_{\rm fus}}{\Delta S_{\rm fus}},
\end{equation}
where $\Delta H_{\rm fus}$ and $\Delta S_{\rm fus}$ are the enthalpy and entropy of fusion. If a \umlp{} captures the shape and curvature of the potential-energy surface within the solid and liquid phases, then the corresponding vibrational and configurational entropy contributions are expected to be approximately preserved. The dominant residual error can then be interpreted as a phase-dependent shift in the absolute potential energy surface, which changes the heat of fusion. Under the approximation
\begin{equation}
\Delta S_{\rm fus}^{\rm PBE} \approx \Delta S_{\rm fus}^{\rm uMLIP},
\end{equation}
the melting temperature can be corrected by rescaling the \umlp{} prediction with the ratio of PBE and \umlp{} heats of fusion,
\begin{equation}
\Tm^{\rm PBE-corr}
\approx
\Tm^{\rm uMLIP}
\frac{\Delta H_{\rm fus}^{\rm PBE}}
     {\Delta H_{\rm fus}^{\rm uMLIP}}.
\label{eq:pbe_correction}
\end{equation}
Here $\Delta H_{\rm fus}^{\rm uMLIP}$ is obtained from the solid and liquid configurations generated by the \umlp{} coexistence workflow, while $\Delta H_{\rm fus}^{\rm PBE}$ is estimated by re-evaluating representative solid-like and liquid-like snapshots with PBE single-point calculations. This correction should be viewed as a first-order enthalpy adjustment rather than an independent PBE coexistence calculation. It is expected to be most reliable when the \umlp{} trajectories remain representative of the corresponding PBE ensembles and less reliable when the \umlp{} samples an incorrect phase, structure, or liquid configuration.

For this reason, both the raw and PBE-corrected results are informative when evaluating a uMLIP. The raw melting temperatures provide the most direct measure of predictive accuracy and reflect the overall ability of the uMLIP to reproduce both the thermodynamics and energetics of melting. In contrast, the PBE-corrected values isolate a different aspect of model quality. If the correction substantially improves agreement with experiment, this suggests that the uMLIP captures the overall shape and curvature of the solid and liquid potential-energy surfaces, and therefore much of the corresponding entropy of the phases, while exhibiting a systematic energy offset that can be corrected through the heat of fusion. Conversely, if the correction does not improve the prediction, then the discrepancy is more likely associated with deficiencies in the sampled solid or liquid configurations, entropy differences, or other transferability limitations that cannot be removed by a simple first-order enthalpy correction. Evaluating both quantities therefore provides complementary information: the raw result measures practical predictive performance, whereas the PBE-corrected result provides insight into how accurately the underlying uMLIP represents the thermodynamic landscape governing melting.

\begin{table*}[t]
\centering
\setlength{\tabcolsep}{2.6pt}
\caption{MeltBench-10 validation results for the three production universal machine-learning interatomic potentials exposed in the deployed \sluschiup{} interface. Entries are melting temperatures in K, with signed errors relative to the reference value shown in parentheses. ``uMLIP'' denotes the direct \sluschiup{} coexistence estimate. ``PBE'' denotes the PBE-corrected value when available; dashes indicate that no PBE correction is available for that entry. Summary rows report the MeltBench aggregate statistics for the raw predictions and for the PBE-corrected set. For the PBE-corrected columns, entries marked with a dagger use the raw \sluschiup{} value because no PBE correction was available for that backend/material pair. Summary statistics in the PBE-corrected block are computed over the displayed values, i.e., corrected values where available and raw fallback values otherwise. Because corrections are available on backend-specific subsets, corrected summary rows should not be interpreted as a strict head-to-head ranking.}
\label{tab:meltbench10}
\begin{tabular}{@{}l r cc cc cc@{}}
\toprule
& & \multicolumn{2}{c}{DPA-3.2-5M-OMat24} &
\multicolumn{2}{c}{Allegro-OAM-L} &
\multicolumn{2}{c}{MACE-MPA-0} \\
\cmidrule(lr){3-4}
\cmidrule(lr){5-6}
\cmidrule(l){7-8}
Material & $T_m^{\rm ref}$ & uMLIP & PBE & uMLIP & PBE & uMLIP & PBE \\
\midrule
Al & 933 & $875\,(-58)$ & \textemdash & $937\,(+4)$ & \textemdash & $824\,(-109)$ & $825\,(-108)$ \\
Cu & 1358 & $1349\,(-9)$ & \textemdash & $1234\,(-124)$ & \textemdash & $1187\,(-171)$ & $1184\,(-174)$ \\
W & 3695 & $3798\,(+103)$ & $3259\,(-436)$ & $3598\,(-97)$ & \textemdash & $3042\,(-653)$ & $3136\,(-559)$ \\
NaCl & 1074 & $1187\,(+114)$ & \textemdash & $1187\,(+114)$ & $1225\,(+152)$ & $1230\,(+157)$ & \textemdash \\
ZrO$_2$ & 2981 & $2997\,(+17)$ & \textemdash & $3042\,(+62)$ & $2867\,(-113)$ & $2518\,(-463)$ & \textemdash \\
La$_2$Zr$_2$O$_7$ & 2553 & $3110\,(+557)$ & $3108\,(+555)$ & $3100\,(+547)$ & $3168\,(+615)$ & $2997\,(+444)$ & \textemdash \\
AlNi$_3$ & 1663 & $2197\,(+534)$ & $1948\,(+285)$ & $2197\,(+534)$ & $1949\,(+286)$ & $2198\,(+535)$ & $1906\,(+243)$ \\
Mo$_{19}$Ru$_{35}$Ta$_2$W$_8$ & 2542 & $2598\,(+56)$ & \textemdash & $2598\,(+56)$ & \textemdash & $2338\,(-204)$ & \textemdash \\
SiC & 3103 & $2800\,(-303)$ & $2677\,(-426)$ & \textemdash & \textemdash & $2599\,(-504)$ & $2967\,(-136)$ \\
HfC & 4228 & $4199\,(-29)$ & $3845\,(-382)$ & $4480\,(+253)$ & $4097\,(-131)$ & $4200\,(-28)$ & \textemdash \\
\midrule
MAE & & 177.8 & 233.6 & 198.8 & 175.3 & 326.6 & 251.4 \\
Mean \% error & & 8.7\% & 9.4\% & 9.6\% & 8.5\% & 14.7\% & 11.5\% \\
RMSE & & 268.0 & 304.3 & 277.4 & 245.1 & 385.4 & 301.6 \\
\bottomrule
\end{tabular}
\end{table*}

\subsection{Compact benchmark results}

Table~\ref{tab:meltbench10} summarizes the compact MeltBench-10 validation results for the three production \umlps{} currently exposed through the public \sluschiup{} interface. For each backend, the table reports the raw coexistence estimate and, where available, the PBE-corrected value defined by Eq.~\eqref{eq:pbe_correction}. Dashes indicate cases for which the PBE correction was not available. This layout separates the direct deployed \sluschiup{} prediction from the optional first-order enthalpy correction. The compact set contains 29 available single-backend raw calculations across ten materials; the missing entry corresponds to an unavailable Allegro-OAM-L result for SiC. Before PBE correction, the mean absolute errors are 177.8~K for DPA-3.2-5M-OMat24, 198.8~K for Allegro-OAM-L, and 326.6~K for MACE-MPA-0, with corresponding root-mean-square errors of 268.0, 277.4, and 385.4~K. With the available PBE corrections applied, the aggregate errors generally improve for Allegro-OAM-L and MACE-MPA-0, whose MAEs decrease to 175.3 and 251.4~K, respectively. The DPA-3.2-5M-OMat24 MAE increases to 233.6~K because several corrected refractory and covalent cases move farther from experiment.

The model-dependent trends in Table~\ref{tab:meltbench10} are also informative. For some systems, such as La$_2$Zr$_2$O$_7$ and AlNi$_3$, all three production \umlps{} significantly overestimate the reference melting temperature, suggesting a shared limitation that may arise from overlapping training data, or common difficulty in representing the relevant high-temperature solid--liquid configurations. In other systems, such as W, ZrO$_2$, and HfC, the three \umlps{} behave quite differently, indicating that model architecture, training distribution, and high-temperature extrapolation can strongly affect melting predictions. These observations support the central design philosophy of \sluschiup{}: model disagreement should be exposed as a diagnostic signal rather than hidden behind a single black-box prediction.

The MeltBench-10 results should be interpreted as infrastructure validation rather than a definitive ranking of universal machine-learning interatomic potentials. The compact benchmark demonstrates that the deployed \sluschiup{} workflow can produce quantitative, atomistic melting-temperature estimates across a diverse set of metals, ionic compounds, oxides, carbides, intermetallics, refractory materials, and a multicomponent alloy. At the same time, the observed model-dependent behavior highlights the importance of uMLIP selection, transferability, and high-temperature sampling. Because the three production backends differ in architecture, training data, and chemical coverage, a rigorous comparison requires a substantially larger benchmark set, consistent correction protocols, uncertainty calibration, and chemistry-resolved analysis. These questions are the focus of the companion MeltBench study. The results presented here therefore support the central claim of this paper: \sluschiup{} provides a functioning, deployed infrastructure for atomistic melting-temperature estimation and validation, rather than a replacement for experiment or high-fidelity first-principles calculations.

\begin{figure*}[t]
\centering
\includegraphics[width=0.9\textwidth]{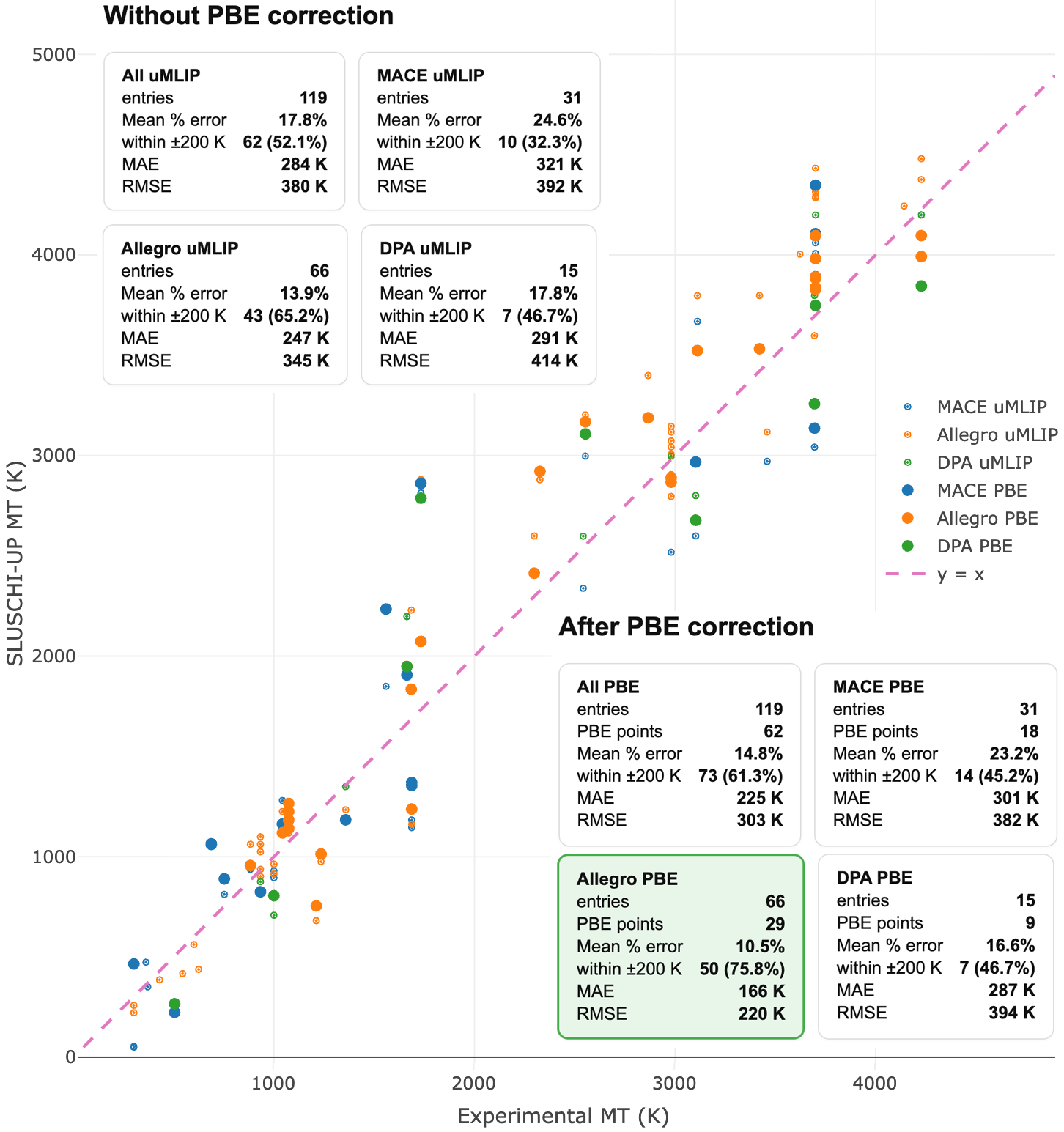}
\caption{Broader MeltBench validation snapshot for the deployed \sluschiup{} workflow. The parity plot compares \sluschiup{} melting-temperature estimates with experimental or reconciled reference melting temperatures. Open symbols denote raw uMLIP coexistence predictions, filled symbols denote PBE-corrected values, and the dashed line indicates perfect agreement, $y=x$. Summary boxes report the number of entries, mean percentage error, number and fraction of predictions within $\pm 200$~K, mean absolute error (MAE), and root-mean-square error (RMSE) for all production uMLIP entries and for each individual backend. The current dataset contains 119 raw uMLIP entries; applying the available PBE corrections reduces the aggregate MAE from 284 to 225~K and the RMSE from 380 to 303~K. The dataset is continuously updated through the public MeltBench platform.}
\label{fig:meltbench_parity}
\end{figure*}

\begin{figure*}[t]
\centering
\includegraphics[width=\textwidth]{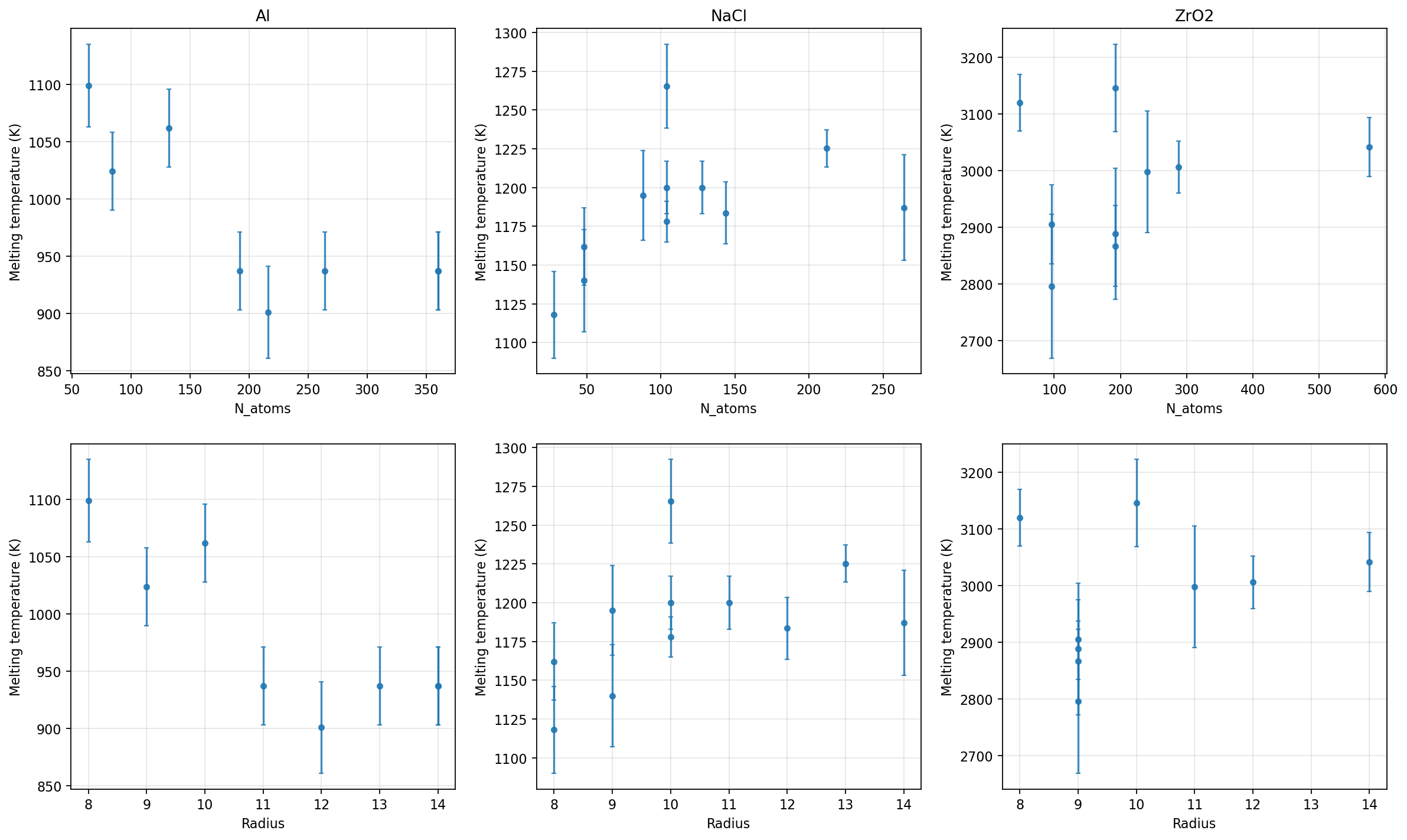}
\caption{Representative finite-size tests in \sluschiup{} using the Allegro-OAM-L backend. The predicted melting temperature is plotted as a function of the number of atoms in the coexistence cell for Al, NaCl, and ZrO$_2$. Error bars denote the statistical uncertainty from the coexistence analysis. The deployed workflow automatically chooses simulation cells with a minimum effective radius of 11~\AA{} \textbf{and} at least 100 atoms when possible. The results indicate that finite-size effects are material dependent: Al shows a stronger size dependence for small cells, whereas NaCl and ZrO$_2$ fluctuate around a comparatively stable range once moderate cell sizes are reached.}
\label{fig:finite_size}
\end{figure*}

\subsection{Broader deployed-job validation}

Beyond the compact MeltBench-10 validation set, the deployed \sluschiup{} service has been tested on a broader and continuously growing collection of benchmark calculations through MeltBench. MeltBench is maintained as the public validation platform associated with \sluschiup{}, where predictions are compared against experimentally measured or reconciled melting temperatures for inorganic materials \cite{meltbenchweb}. The dataset includes raw \sluschiup{} coexistence estimates, optional PBE-corrected values, statistical uncertainties, simulation-cell sizes, effective radii, and reference melting temperatures. The public MeltBench page also provides the benchmark data and column definitions, allowing the dataset to be downloaded and reanalyzed as the benchmark grows \cite{meltbenchweb}.

Figure~\ref{fig:meltbench_parity} summarizes the current deployed-job validation set as a parity plot of \sluschiup{} melting temperatures against experimental or reconciled reference values. Without PBE correction, the current dataset contains 119 uMLIP entries, with an overall mean absolute error of 284~K, root-mean-square error of 380~K, and 62 predictions within $\pm 200$~K of reference, corresponding to 52.1\% of entries. Among the three production backends, Allegro-OAM-L gives the lowest raw MAE in this broader set, 247~K over 66 entries, followed by DPA-3.2-5M-OMat24 with 291~K over 15 entries and MACE-MPA-0 with 321~K over 31 entries. These statistics are reported here as a deployment-scale snapshot rather than a definitive model ranking, because the dataset is still growing and the materials distribution is not yet balanced across backends.

The PBE correction generally improves the aggregate agreement with reference melting temperatures. Using the current PBE-corrected dataset, the overall MAE decreases from 284 to 225~K and the RMSE decreases from 380 to 303~K, while the fraction of predictions within $\pm 200$~K increases from 52.1\% to 61.3\%. The improvement is especially clear for Allegro-OAM-L, for which the MAE decreases from 247 to 166~K and the fraction within $\pm 200$~K increases from 65.2\% to 75.8\%. The correction is not uniformly beneficial for every backend or material, which is expected because it is a first-order enthalpy correction rather than a full first-principles coexistence recalculation. Nevertheless, the broader deployed-job statistics support the central conclusion that \sluschiup{} provides a practical screening-level workflow, while the detailed chemistry-resolved interpretation and full leaderboard analysis are reserved for the companion MeltBench paper.

\section{Discussion}

Finite-size effects are an important practical consideration for a deployed small-cell coexistence workflow. In \sluschiup{}, the simulation cell is automatically selected to satisfy a minimum effective radius of 11~\AA{} \textbf{and} at least 100 atoms whenever possible. Figure~\ref{fig:finite_size} shows representative size-dependence tests for Al, NaCl, and ZrO$_2$ using the Allegro-OAM-L backend. The results show the system-dependent behavior: small cells can introduce noticeable shifts, while larger cells generally produce estimates that fluctuate within the statistical uncertainty of the coexistence fit. These observations are consistent with the original DFT-based SLUSCHI studies, which motivated the use of simulation cells containing approximately 100 atoms or more as a practical balance between finite-size accuracy and computational efficiency \cite{hong2013small,hong2016sluschi}.

Table~\ref{tab:baselines} places \sluschiup{} in the broader landscape of melting-temperature prediction and materials cyberinfrastructure. Direct graph-neural-network predictors provide rapid scalar estimates but do not generate atomistic coexistence trajectories. First-principles \sluschi{} provides a physically grounded coexistence workflow but remains too expensive for routine deployment. 
\sluschiup{} should be viewed as an atomistic melting-temperature screening and validation platform rather than as either a direct scalar predictor or a replacement for high-fidelity first-principles simulation. By combining \sluschi{}-style small-size coexistence with universal machine-learning interatomic potentials, the platform occupies a middle ground between graph-neural-network melting-temperature predictors, which provide rapid estimates but limited physical diagnostics, and first-principles coexistence calculations, which provide stronger physical grounding but remain computationally expensive. The deployed workflow also exposes model disagreement and records job-level provenance, allowing the same structure to be evaluated with different uMLIPs and making predictions reproducible across model versions and checkpoints. At the same time, several limitations remain important. The current production backends are not universally reliable across all chemistries, phases, pressures, and liquid configurations; melting calculations probe high-temperature and interfacial environments that are often outside the most densely sampled regions of uMLIP training data.

Recent work accelerated \sluschi{} by coupling the workflow to LAMMPS and pre-trained LASP neural-network potentials, demonstrating computational-cost reductions exceeding one order of magnitude relative to conventional DFT simulations across a benchmark set of 30 material systems \cite{campbell2026sluschi}. \sluschiup{} builds on this direction but addresses a different problem: deployment, accessibility, and generalizability. Rather than focusing on a single acceleration study, \sluschiup{} provides a public web interface through which users can submit structures, select among multiple universal machine-learning potentials, and run \sluschi{}-style coexistence calculations through a shared GPU queue. The contribution of \sluschiup{} is therefore infrastructural as well as methodological, providing a practical route for \sluschi{}-based melting calculations without requiring users to maintain local molecular-dynamics workflows.

The need for caution is consistent with recent assessments of universal machine-learning interatomic potentials. Out-of-the-box uMLIPs can show systematic softening and reduced accuracy for high-energy, defect, surface, migration, and other out-of-distribution configurations \cite{deng2024softening}. Surface-energy benchmarks have likewise shown that broadly trained universal models can perform poorly when applied to environments far from bulk near-equilibrium training data \cite{focassio2024surfaces}. Because melting calculations involve liquid-like, interfacial, and high-temperature configurations, \sluschiup{} treats model disagreement and trajectory diagnostics as part of the prediction rather than as incidental metadata.

Beyond the technical implementation, \sluschiup{} is intended to lower the barrier to atomistic melting-temperature calculations. Historically, coexistence-based melting studies have required specialized software, substantial computational resources, and significant expertise in molecular dynamics workflows. By exposing a validated \sluschi{} workflow through a public web service and allowing one free calculation per verified user every 24 hours, \sluschiup{} enables broader participation in melting-temperature prediction and validation. Coupled with the growing MeltBench benchmark platform, this accessibility may help establish more reproducible and transparent standards for evaluating both melting-temperature predictions and universal machine-learning interatomic potentials.

\begin{table*}[t]
\centering\small
\caption{Methods relevant to evaluating \sluschiup{}.}
\label{tab:baselines}
\begin{tabular}{@{}lll@{}}
\toprule
\tcell{0.20\textwidth}{Method} &
\tcell{0.36\textwidth}{Strength} &
\tcell{0.34\textwidth}{Limitation / role} \\
\midrule

\tcell{0.20\textwidth}{First-principles \sluschi{}} &
\tcell{0.36\textwidth}{Automated small-size coexistence with DFT molecular dynamics; strong physical grounding \cite{hong2013small,hong2016sluschi}.} &
\tcell{0.34\textwidth}{Computationally expensive; reference method for selected cases.} \\[0.6em]

\tcell{0.20\textwidth}{Formula-based GNN predictor} &
\tcell{0.36\textwidth}{Very fast melting-temperature estimates trained on large melting-temperature data sets \cite{hong2022gnn}.} &
\tcell{0.34\textwidth}{Does not produce atomistic coexistence trajectories; limited pressure and mechanistic diagnostics.} \\[0.6em]

\tcell{0.20\textwidth}{\sluschi{} + LAMMPS + LASP} &
\tcell{0.36\textwidth}{Demonstrated machine-learning-potential acceleration of the \sluschi{} workflow with substantial speedups \cite{campbell2026sluschi}.} &
\tcell{0.34\textwidth}{Depends on availability and transferability of selected potentials.} \\[0.6em]

\tcell{0.20\textwidth}{\sluschiup{}} &
\tcell{0.36\textwidth}{Web-deployed, provenance-tracked, uMLIP-accelerated coexistence workflow.} &
\tcell{0.34\textwidth}{Validated here on MeltBench-10 as an infrastructure demonstration; full model ranking and uncertainty calibration are reserved for MeltBench.} \\

\bottomrule
\end{tabular}
\end{table*}

\section{Future Work}

The current production deployment exposes three validated uMLIP backends, while the beta interface already includes additional experimental models such as PET, SevenNet, and CHGNet. Future work will focus on evaluating these and other emerging universal machine-learning interatomic potentials through the MeltBench framework and expanding the range of materials available for public validation. As the benchmark grows, it will provide a broader picture of uMLIP performance for melting-temperature prediction across different chemistries and materials classes.

More broadly, \sluschiup{} and MeltBench are intended to evolve together: \sluschiup{} provides a public platform for atomistic melting-temperature calculations, while MeltBench provides a continuously expanding benchmark for assessing the strengths and limitations of universal machine-learning interatomic potentials. Continued development will focus on improving reliability, expanding benchmark coverage, and strengthening reproducibility and transparency in melting-temperature prediction.

\section{Conclusion}
I presented \sluschiup{}, a deployed web infrastructure for melting-temperature estimation using \sluschi{}-style small-size solid–liquid coexistence simulations accelerated by universal machine-learning interatomic potentials. The current public service accepts Materials Project identifiers or POSCAR structures, automatically constructs coexistence simulations, executes molecular dynamics through a shared GPU queue, and returns atomistic melting-temperature estimates without requiring users to install simulation software or maintain local high-performance-computing resources. The production deployment currently supports \texttt{mace-mpa-0-medium}, \texttt{Allegro-OAM-L}, and \texttt{DPA-3.2-5M-OMat24}, while additional experimental backends are available through the beta interface.

Validation on the compact MeltBench-10 benchmark demonstrates that the deployed workflow can produce useful screening-level melting-temperature estimates across metals, ionic solids, oxides, carbides, intermetallics, refractory materials, and multicomponent alloys. The broader MeltBench validation dataset further shows that atomistic melting calculations can now be performed reproducibly and at scale through a public web platform. Although the reported accuracies depend on the selected uMLIP, finite-size effects, and high-temperature model transferability, the results demonstrate that modern universal interatomic potentials are sufficiently mature to support practical coexistence-based melting calculations across a wide range of materials.

The principal contribution of \sluschiup{} is therefore not a new melting algorithm, but a reproducible and publicly accessible infrastructure layer for atomistic melting-temperature prediction. To our knowledge, SLUSCHI-UP is the first publicly accessible web service dedicated to atomistic melting-temperature prediction using universal machine-learning interatomic potentials. By lowering the barrier to advanced coexistence simulations and coupling them with an open benchmarking framework, \sluschiup{} provides a foundation for broader community validation, comparison, and improvement of universal machine-learning interatomic potentials. More broadly, the platform has the potential to serve as a common reference workflow for melting-temperature calculations, making atomistic melting simulations more accessible, repeatable, and transparent to the materials-science community.

\section*{Acknowledgements}

The authors acknowledge Arizona State University Research Computing for hosting the deployed \sluschiup{} service. This work used Delta GPU at NCSA through allocation MAT260050 from the Advanced Cyberinfrastructure Coordination Ecosystem: Services \& Support (ACCESS) program, which is supported by U.S. National Science Foundation grants \#2138259, \#2138286, \#2138307, \#2137603, and \#2138296.

\section*{Data and Code Availability}

The deployed \sluschiup{} service is available at \url{https://faculty.engineering.asu.edu/hong/sluschi-up/}. 
The MeltBench validation page is available at \url{https://jobs.sluschi-mapp.org/meltbench}. 
The original DFT-based \sluschi{} code is available through the public GitHub repository cited in Ref.~\cite{sluschigithub}. 

\section*{Competing Interests}

The authors declare no competing interests.

\bibliographystyle{apsrev4-2}
\bibliography{sluschi-up}

\end{document}